\documentclass[12pt]{article}
\usepackage{fleqn,amssymb,amsfonts}
\usepackage{latexsym}

\newcommand {\bR}{{\Bbb R}}
\newcommand {\bN}{{\Bbb N}}
\newcommand {\bZ}{{\Bbb Z}}
\newcommand {\bI}{{\Bbb I}}
\newcommand {\bC}{{\Bbb C}}

\newcommand {\bQ}{{\Bbb Q}}
\newcommand {\bpsi}{{\bf \psi}}
\newcommand {\bphi}{{\bf \phi}}
\newcommand {\cB}{{\cal B}}

\newcommand {\cO}{{\cal O}}

\newcommand {\cT}{{\cal T}}
\newcommand{\beq}{\begin{equation}}
\newcommand{\eeq}{\end{equation}}
\newcommand{\beqn}{\begin{eqnarray}}
\newcommand{\eeqn}{\end{eqnarray}}
\newcommand{\beqno}{\begin{eqnarray*}}
\newcommand{\eeqno}{\end{eqnarray*}}
\newtheorem{theorem}{Theorem} [section]
\newtheorem{lemma}[theorem]{Lemma}

\newtheorem{corollary}[theorem]{Corollary}

\newtheorem {remark}[theorem]{Remark}
\newtheorem {remarks}[theorem]{Remarks}

\begin{document}
\title {Stability of driven  systems with growing gaps,\\
        Quantum rings and Wannier ladders}
\author{Joachim Asch${}^\ast$\and Pierre Duclos\thanks{CPT-CNRS,
Luminy Case 907,  F-13288 Marseille Cedex 9, France and PhyMat,
Universit{\'e} de Toulon et du Var, BP 132, F-83957  La Garde Cedex.
e-mail: asch@cpt.univ-mrs.fr, duclos@naxos.unice.fr}
\and  Pavel Exner
\thanks{Nuclear Physics Institute, Academy of Sciences, CZ--25068 \v
Re\v z, and  Doppler Institute, Czech Technical University,
CZ--11519 Prague, Czech Republic. email:  exner@ujf.cas.cz} }
\date{26/5/98 }
\maketitle
\begin{abstract} We consider  a quantum particle in a periodic
structure submitted to a constant external electromotive force. The
periodic background is given by a smooth potential plus  singular
point interactions and has the property that the gaps between its bands
are growing with the band index. We prove that the
spectrum is pure point--i.e. trajectories of wave packets lie in
compact sets in Hilbert space-- if the Bloch frequency is non-resonant
with the frequency of the system and satisfies a Diophantine
type estimate, or if it is resonant. Furthermore it
is shown that  the KAM method employed in the non-resonant case
produces uniform bounds on the growth of energy for driven systems. 

 \end{abstract} %

 \section{Introduction}
 We study stability of the dynamics  of one electron in a 1d
periodic structure with infinitely many open gaps driven by a constant
electromotive force. To be specific we
consider two realizations: the Stark-Wannier problem for
a periodic background  interaction $V(x)=V(x+L)$ defined by the
Hamiltonian
\beqno H_S=-\Delta+V(x)+Fx\qquad{\hbox{on } L^2({\bR})}\eeqno  and
an electron on a conducting ring in the plane
threaded  by a linearly increasing magnetic flux line
$\Phi(t)=Ft$ whose dynamics is defined
by
\beqno H_R(t)=(-i\partial_x-Ft)^2+V(x)\qquad{\hbox{on }
L^2({S^1})}.\eeqno  

The stability of a general time dependent system  was
addressed for example in
\cite{enssvese,buni+4,deol,nenc,joye,debiforn,barbjoye}. The
discussion  of stability may be summarized in the question whether a
wave packet can get delocalized during its time evolution. To answer
this  one may study the time behavior of expectations
of observables; if the system is periodic in time  the
spectral properties of the Floquet operator--i.e.: the evolution over
one period-- can provide precise information on the stability. To
mention one example: if the periodically time dependent system is
confined and unbounded and the spectrum of its Floquet operator is
absolutely continuous then the energy expectation grows in time for
any initial state, so the system is unstable.  

The special case considered here was  intensively studied since
Wannier conjectured existence of ladders of eigenvalues; see
\cite{grecsacc,bastmend} for background on this story.  In
\cite{6} it was proven that for smooth background potential $V\in
C^2(\bR)$ --in fact $C^{1+\epsilon}$-- the spectrum of $H_S$ is
absolutely continuous which leads to unbounded growth of the energy
for $H_R(t)$, see \cite{avronemi}. On the other hand in
\cite{avroexnelast}, \cite{exne}  a comb of
$\delta^\prime$ point interactions was considered. It was shown
that this model is physically important,  in particular it describes
idealized geometric scatterers. It was proven that the
spectrum has no absolutely continuous component leaving the
possibility of eigenstates, singular continuous spectrum and
unbounded energy growth. Furthermore  a conjecture on the
essential spectrum was made. See also
\cite{kise} for the geometric scatterer aspect and
\cite{maiosaccerratum} for a second proof of absence of absolute
continuity.

 It was argued by Ao \cite{ao} that the spectral nature depends on
the gap structure of the periodic background. He conjectured that
for gap behavior $\Delta E_n=\cO(1/n^\alpha)$ one has point
spectrum for
$\alpha<0$ at least for ``non-resonant" $F$ and continuous spectrum
for
$\alpha>0$. For $\alpha=0$ a phase transition from pure
point to continuous spectrum with growing $F$ is expected
--see also
\cite{bereovch}, \cite{bentogrecziro}; furthermore the spectral
nature  seems to depend also on number theoretical
properties of the driving frequency $FL$.
This critical case corresponds to the
driven Kronig-Penney model; another realization of constant gaps is
the --explicitly solvable-- forced harmonic oscillator considered by
\cite{enssvese,hagelossslaw,comb,buni+4}. 

Our contribution, here, is to show stability for $V$ a comb of
$\delta^\prime$ interactions plus a smooth bounded background.
We prove that the spectrum of $H_S$ is pure point;
$H_R$ is periodic up to a gauge transformation, the
Floquet Hamiltonian of the transformed problem is unitarily
equivalent to $H_S$, its spectrum is also pure point. 
 We are able to prove this
in two different settings. Firstly for a large subset  of frequencies
$FL$  which do not resonate with the frequencies
 of the background we use
a KAM algorithm in order to treat the $\delta^\prime$ interaction as
perturbation of the decoupled problem where $\delta^\prime$ is
replaced by a Neumann boundary condition. This algorithm needs as
input a matrix which has sufficient off diagonal decay. Because of the
singularity of the interaction it is not evident that the Floquet
Hamiltonian considered here has such a matrix representation.
A detailed spectral analysis is necessary
to show that this is the case. Technically the  basic observation  is
that the eigenfunctions do not concentrate at the singularity if the
band index grows. A consequence is that the gaps are increasing and
the transition matrix has the required properties. This result was
announced in \cite{aschduclexne1}. In the second case for the
countable set of resonant frequencies we prove the conjecture of
\cite{avroexnelast,exne} concerning the location of the essential
spectrum; a general argument based again on the off-diagonal decay
of matrix elements allows to conclude that also in this case the
spectrum is pure point.

These results strongly suggest that in
the models considered here, in fact for $\alpha<0$ in the Ao language
and for reasonable boundedness of the transition matrix, pure
pointness of the spectrum should not depend on number theoretical
properties of the frequency!

The fact that the spectrum of the Floquet operator is pure point--
in the $H_R$ picture-- does not imply on general grounds that 
the energy expectation  is bounded in time 
as the example in \cite{delrjitolastsimo2} shows. We prove here that
applicability of the KAM method provides a uniform bound on the
energy growth, so this applies here to the non-resonant case and holds
true for a subclass of general time dependent Hamiltonians studied in
\cite{nenc,joye} which complements their results.

The organization of the paper is as follows. In section
\ref{matrix} we fix notations, define the problem in detail and
provide a regularization
necessary for the methods in the following sections to work.
Section \ref{nonresonant} is devoted to the study of the non-resonant
case; the result on pure pointness is Corollary \ref{purepoint}, the
bounds on energy growth are Corollary
\ref{energygrowth} and Theorem \ref{generalenergygrowth}. In section
\ref{resonant} we determine the essential spectrum and prove
pure pointness in the resonant case.

\section{The problem and its Matrix Representation}\label{matrix} We
consider the class of potentials
\beqno V(x)=\sum_{n\in\bZ}\beta\delta^\prime(x-nL)+W(x)\eeqno  with
$W(x)=W(x+L)$ a differentiable multiplication operator on $L^2(\bR)$
and
$\delta^\prime$ defined by the expressions (\ref{deltaprime}) below.
We refer to \cite{avroexnelast} and references therein for
background material on this model.

The limit $\beta\to\infty$ represents decoupling of the cells by a
Neumann boundary condition. For $\beta$ large the problem is a
perturbation of the decoupled case, but only in the
quadratic form sense. We shall show in this chapter that in spite of
the singular character of the perturbation the problem can be
represented by a matrix operator with polynomial off-diagonal decay.
Let us fix 

{\bf Notations}.  $L,F,\beta$ are positive numbers. We shall,
however, mostly work with  the parameters $\omega:=FL, g:=1/\beta$
and employ the symbols $T:={2\pi\over FL}$ for the Bloch period;
$D:=-i\partial$ where $\partial$ denotes a partial derivative; ${\bf
D}M({\bf O}M)$ for the diagonal (off-diagonal) part of a  matrix $M$;
$\chi$ for the binary code defined by $\chi(True):=1, \chi(False):=0$;
$cte$ for a generic constant, independent of the parameters, which may
change from line to line. $H^n$ denotes a Sobolev space of order $n$.
We shall try to avoid to note the dependence of parameters of a
quantity if we feel that this is possible while keeping clarity.

The Stark-Wannier Hamiltonian is
\beqno H_S=D_x^2+Fx+W(x)\eeqno  defined on 
\beqn &&D(H_S)=\{\psi\in H^2(\bR\setminus(\bZ L)), H_S\psi\in
L^2(\bR),\nonumber\\ 
&&\psi^\prime(nL+)-\psi^\prime(nL-)=0,\psi(nL+)-\psi(nL-)=
\beta\psi^\prime(nL)\}.\label{deltaprime}\eeqn 

The Hamiltonian for the driven ring is informally :
\beqno H_R(t)=\nabla^2+W(x)+(\nabla\delta)\qquad\hbox{on }
L^2(S^1),\eeqno 
where  $\nabla:=D_x-Ft$;
we use, however,  the time dependent gauge transformation
$\exp{(-iFtx)}$ to transform the propagator into the periodic one of
period $T$ generated by the Hamiltonian
\beqn\label{H(t)} H(t)&=&H(t,\omega,g):=D_x^2+{\omega\over
L}x+W(x)\\  D(H(t))&:=&\{\psi\in H^2((0,L); e^{i\omega
t}\psi^\prime(L)-\psi^\prime(0)=0,\nonumber\\
&& g(e^{i\omega t}\psi(L)-\psi(0))=-\psi^\prime(0)\}.\nonumber\eeqn

Notice that because the domain is $t$ dependent an argument
for existence of the propagator is needed. This will be shown by
mapping the problem to one whose propagator is known to exist, see
 remark \ref{existenceofpropagator} at the end of this section.

 We shall henceforth study
the point spectrum of the Floquet Hamiltonian 
\beq\label{K} K=K(\omega,g)=D_t^{per}+H(t,\omega,g)\eeq 
acting in
$L^2((0,T),dt;L^2((0,L),dx))$  on the domain
\beqno D(K)=\{\bpsi\in
H^1((0,T),D(H(t,\omega,g)),\bpsi(T,x)=\bpsi(0,x))\}.\eeqno

An eigenvector $\phi$ of $K$ with eigenvalue $\epsilon$ will provide
us with a Bloch-Floquet solution of the Schr\"odinger equation
\beqno D_t\bpsi(t,x)+H(t)\bpsi(t,x)=0\eeqno which is of the form
\beqno \bpsi(t,x)=e^{-i\epsilon t}\bphi(t,x)\eeqno
with $\bphi$ periodic in $t$.

A second reason to introduce $K$ is the unitary equivalence of $H_S$
and
$K$, see \cite{avronemi}:
\beqno U_BH_S=KU_B\eeqno where  $U_B$ is the
Bloch transformation

\beqno U_B:L^2(\bR)\rightarrow L^2((0,T)\times(0,L), dt\ dx)\\
       (U_B\psi)(t,x)={1\over\sqrt{T}}\sum_{\gamma\in\bZ}
e^{i\gamma \omega t}\psi(x+\gamma L).\eeqno

The matrix representation $M$ of $K$ is constructed as follows:
let $\{\psi_n(t)\}_{n\in\bN}$ be a periodic orthonormal eigenbasis of
$H(t)$: $\psi_n(t+T)=\psi_n(t)$. 
\beqno\{\phi_j\}_{j\in\bZ\times\bN}, \qquad\phi_j(t,x)=
\phi_{j_1,j_2}(t,x):={e^{i\omega t j_1}\over\sqrt{T}}
\psi_{j_2}(t,x)\eeqno then is a basis of
$L^2((0,T),dt;L^2((0,L),dx))$. Using
$\langle., .\rangle$ for  the scalar product in $x$ space we define
\beqn\label{M}
{M}_{jk}&=&{\bf\langle\langle}\phi_j,K\phi_k{\bf\rangle\rangle}:=
\int_0^T\langle \phi_j,K\phi_k\rangle(t)\ dt\\ &=&{1\over
T}\int_0^T e^{i(j_1-k_1)\omega t}
\left((k_1\omega+E_{k_2}(t))\delta_{j_2k_2}+
\langle\psi_{j_2},D_t\psi_{k_2}\rangle(t)\right)\ dt.\nonumber\eeqn

In the rest of this section we shall study the properties of the
eigenvalues
$E_n(t,\omega,g)$ of $H$ and the coupling matrix
$\langle\psi_n, D_t\psi_m\rangle$. 

For $\psi\in H^1((0,L)), \phi\in D(H(t))$ we find by integration by
parts:
\beqno\langle\psi,
H\phi\rangle&=&\langle\psi^\prime,\phi^\prime\rangle+\langle\psi,
({\omega\over L}x+W)\phi\rangle\\ &&+g \overline{(e^{i\omega
t}\psi(L)-\psi(0))}(e^{i\omega t}\phi(L)-\phi(0)).\eeqno

So denoting the Neumann decoupled operator ($g=0$) by
\beqno H_0=D_x^2+{\omega\over L} x + W(x)\qquad\hbox{with }
\psi^\prime(0)=\psi^\prime(L)=0\eeqno we have the representation
\beqno\label{hkrein} H=H_0+g\vert f(t,\omega)\rangle\langle
f(t,\omega)\vert\eeqno  where
\beqno f(t,\omega)=e^{-i\omega t}\delta_L-\delta_0.\eeqno
$f$ is in $H^{-1}((0,L))$ so we are in the framework of generalized
rank-one perturbations; we shall use the results of
\cite{simocanada}.  
$ H(t,\omega,g)$ is an analytic family with constant form domain
$H^1(0,L)$  for $(t,\omega,g)\in S_{\alpha_t}\times
S_{\alpha_\omega}\times\bC$ for some
${\alpha_t},{\alpha_\omega}>0$ where
$S_\alpha:=\{z\in\bC;\vert
\hbox{Im} z\vert<\alpha\}$. For the resolvent $R(z)=(H-z)^{-1}$ it
holds:
\beq\label{krein} R(z)-R_0(z)=-{g\over 1+g G(z)}\vert
R_0(z)f\rangle\langle R_0(\bar z)f\vert\eeq
 with $G(z)=G(z,t,\omega):=\langle f,R_0(z)
f\rangle$. 

In the sequel we make statements for $g$ small enough. This could be
circumvented by the use of an adiabatic technique. We shall not do
so as in section \ref{nonresonant} the smallness of $g$ will be
essential anyhow. We obtain

\begin{theorem}\label{spectrum} For $g$ sufficiently small,
$\omega$ in a
given interval $\lbrack\omega_-,\omega_+\rbrack\subset(0,\infty)$, 
$T={2\pi\over\omega}$,
$t\in\lbrack0,T\rbrack$ the operator $H(t,\omega,g)$ as defined
in equation (\ref{H(t)}) has simple discrete spectrum. For its
eigenvalues
$E_n=E_n(t,\omega,g)$ it holds uniformly in
$t,\omega,g,n$:
 \begin{enumerate}
\item[(i)] ${E_{n+1}-E_n\over n}\ge cte>0,$
\item[(ii)] $0\le\partial_\omega E_n\le cte<1;$
\end{enumerate}
 furthermore there exists a  basis 
 $\{\psi_n\}$ of eigenfunctions of $H$ with
$\psi_n(t+T)=\psi_n(t)$, $\psi_n\in
C^\infty(\lbrack0,T\rbrack\times\lbrack\omega_-,\omega_+\rbrack
\times\lbrack0,g_{max}\rbrack)$ such that in the $C^\infty$ topology
and uniformly in $t,\omega,n$:
\begin{enumerate}
\item[(iii)]$\langle\psi_n,D_t\psi_m\rangle={O(g)\over
\vert n^2-m^2\vert}\qquad \hbox{for } n\neq m ,$ 
\item[(iv)] $\langle \psi_n,D_t\psi_n\rangle=O(g/n).$
\end{enumerate} 
\end{theorem}

{\bf Proof.} 
The behavior for large $n$ has to be controlled. We compare $H$ to
the Neumann Laplacian
$-\Delta^N$  on
$H^2(0,L)$ with boundary conditions
$\psi^\prime(0)=\psi^\prime(L)=0$, whose eigenvalues are $(\pi
n/L)^2$. This is done in two steps: first we compare
$H(t,\omega,0)$ --which is actually time independent-- to
$-\Delta^N$ using regular perturbation theory; secondly the
difference
$H(g)-H(g=0)$ is treated using formula (\ref{krein}).

By a Wronskian argument the eigenvalues of $H(t,\omega,0)$ are
simple and for $n$ large enough it holds: 
\beqno\vert E_n (t,\omega,0)- ({\pi n\over L})^2\vert\le cte\Vert
{\omega\over L}x+W\Vert.\eeqno
The reason why the transition matrix decays and eigenvalues stay nearby
upon switching on
$g$ is contained in the following auxiliary result.

\begin{lemma}\label{f}
For $\phi\in L^2((0,L))$ and the eigenprojections $P_n$ of $H$ it
holds in the $C^\infty$ topology
and uniformly in $n$:
\beqno \vert (P_n(t,\omega,g)\phi)(L)\vert+\vert
(P_n(t,\omega,g)\phi)(0)\vert\le cte
\Vert\phi\Vert\eeqno
\end{lemma}

{\bf Proof.}(Of the Lemma)
By Riesz's formula we have with a circle $\Gamma_n$ of length
$\vert\Gamma_n\vert$ centered at $(\pi n/L)^2$ 
\beqno P_n=-{1\over2\pi
i}\oint_{\Gamma_n}R(z)\ dz.\eeqno
Denote be $a,b$ indices which take the values $0$ and $L$. In
order to prove the estimate on
$P_n\phi(a)=\langle\delta_a,P_n\phi\rangle$ by Krein's
formula--equation (\ref{krein})--  it is sufficient to show
\beqno\sup_{z\in\Gamma_n}(\vert\langle\delta_a,R_0(z)\phi\rangle\vert+
\vert\langle\delta_a,R_0(z)\delta_b\rangle\vert)=
O({1\over\vert\Gamma_n\vert}).\eeqno
To do this we use regular perturbation theory and the fact that the
"gaps" of the Neumann operator are growing. Denote 
${d_n}:={\pi^2\over
L^2}((n)^2-(n-1)^2)$ and choose $n_0$ such that 
$\sup_\omega\Vert{\omega\over L}x+W\Vert<{d_{n_0}\over2}$; for
$n>n_0$ choose a suitable number $M$ and   and
$\vert\Gamma_n\vert:={d_n\over M}$.
 Then it holds for the resolvent $R^N$ of the Neumann Laplacian
\beqno\Vert({\omega\over
L}x+W)R^N(z)\Vert\le{cte\over n}\le1\hbox{ and }
\Vert R^N(z)\Vert=O({1\over n})\eeqno
uniformly for $z\in\Gamma_n$ and

\beqno R_0&=&R^N(1+({\omega\over
L}x+W)R^N)^{-1}\\ 
&=&R^N-R^N(1+({\omega\over
L}x+W)R^N)^{-1}({\omega\over
L}x+W)R^N.\eeqno

So the question is reduced to the explicit calculation of 
$\Vert\delta_a R^N\Vert$ and $\langle\delta_a,
R^N\delta_b\rangle$. $R^N(z)$ is given by its kernel
$R^N(z)\phi(x)=\int R^N(x,y;z)\phi(y)\ dy$
\beqno R^N(x,y;z):=-{1\over \sqrt{z}\sin{\sqrt{z}L}}
\cos(\sqrt{z}(x\land y))\cos(\sqrt{z}((y\lor x)-L))\eeqno
with the notation $x\land y$ ($x\lor y$)for the minimum (maximum) of
$x$ and $y$. One finds for example
\beqno \Vert\delta_0 R^N(z)\Vert\le\Vert
{1\over \sqrt{z}\sin{\sqrt{z}L}}\cos{\sqrt{z}(y-L)}\Vert=
O({1\over n})\eeqno
and similarly $\Vert\delta_L R^N(z)\Vert=O({1\over n}), 
\vert\langle\delta_a, R^N(z)\delta_b\rangle\vert=O({1\over n})$ for
$a,b$ in $\lbrace0,L\rbrace$ uniformly for $z\in\Gamma_n$. These
estimates are preserved upon differentiation with respect to $t,
\omega, g$.\hfill $\Box$\\[2mm]

To continue with the proof of the theorem for the eigenvalues we
show
\beq\label{en} E_n(t,\omega,g)=({n\pi\over
L})^2+{\omega\over2}+\langle W\rangle+{4g\over L}(1-(-1)^n\cos{\omega
t})+O(1/n)\eeq 
in the $C^\infty$ topology. $\langle
W\rangle$ denotes the mean value ${(1/L)}\int_0^LW$. Indeed by regular
perturbation theory with the notation
$V:={\omega x\over L}+W+g\vert f\rangle\langle f\vert$ it is a
corollary of the previous lemma that
\beqno &&E_n-({n\pi\over L})^2-tr(P_n^NV)\\
&&=-{1\over2\pi
i}tr(\oint_{\Gamma_n}(z-({n\pi\over L})^2)R(z)VR^N(z)VR^N(z)\
dz)=O(1/n)\eeqno 
With $P_n^N={2\over L}\vert\cos{{n\pi\over L}x}\rangle\langle
\cos{{n\pi\over L}x}\vert\quad (n\ge1)$ the explicit term of the
approximation follows. From this we infer the assertions concerning
the eigenvalues for $n$ large enough. For
the lowest finitely many $n$ we employ continuity of
$\partial_\omega E_n$ and compactness of
$\lbrack\omega_-,\omega_+\rbrack$ to deduce (ii). 

The eigenfunctions are now constructed as
\beqno
\psi_n(t,\omega,g):={P_n(t,\omega,g)\psi_n^0(\omega)\over\Vert
P_n(t,\omega,g)\psi_n^0(\omega)\Vert}\eeqno
for any time independent choice of eigenfunctions $\psi_n^0$ of
$H_0$. 
We differentiate $HP=EP$ in the quadratic form sense to get 
$P_m\partial_tP_n={P_m\partial_tHP_n\over E_n-E_m}$. It follows
from Lemma \ref{f} for the off-diagonal part:
\beqno
\langle\psi_m,D_t\psi_n\rangle={\langle\psi_m,D_tH\psi_n\rangle\over
E_n-E_m}={O(g)\over \vert n^2-m^2\vert}.\eeqno
For the diagonal a calculation using Lemma \ref{f} yields
\beqno\langle\psi_n,D_t\psi_n\rangle={1\over\Vert\psi_n\Vert^2}
{1\over2}\langle\psi_n^0,\lbrack
P_n,D_tP_n\rbrack\psi_n^0\rangle=O({g\over n}).\eeqno
\hfill
$\Box$\\[2mm]

\begin{remark}\label{existenceofpropagator}
The existence of the propagator $U(t,s)$  of $H(t)$ is a corollary of
the preceding theorem: denote by $J(t)$ the unitary between $l^2(\bN)$
and $L^2((0,L))$ which maps the n'th canonical base vector to
$\psi_n(t)$. Then $J^{-1}(t)(D_t+H(t))J(t)=D_t+h(t)$ where the matrix
operator $h$ is defined by
\beqno
h_{nm}=E_n\delta_{nm}+\langle\psi_n,D_t\psi_m\rangle.\eeqno $h$
is analytic with constant domain so its propagator $u(t,s)$ exists.
$U$ is then given by \beqno U(t,s)=J(t)u(t,s)J^{-1}(s).\eeqno
\end{remark}

\section{Stability for non-resonant frequencies}\label{nonresonant}
In this section we shall employ the KAM algorithm to diagonalize the
matrix $M$ of the Floquet operator $K$. $M$ is considered as a
perturbation of its diagonal ${\bf D}M$. For generic values of the
frequency $\omega$ the eigenvalues of ${\bf D}M$ form a dense subset
of the real line \cite{duclstovvitt}.  We shall show in Corollary
\ref{purepoint} that for a large set of ``non-resonant" $\omega$ the
spectrum of $K$ is pure point, in Corollary \ref{energygrowth} and
in Theorem
\ref{generalenergygrowth} that the energy of the system stays
bounded.

In order to measure the decay of matrix elements consider the
following Banach algebras (see \cite{duclstov}): let $r,\delta\ge0,
\Omega\subset(0,\infty), \langle x\rangle:=(1+x^2)^{1/2}$,
\beqno&&\cB(\Omega,r,\delta):=\left\{\right.\omega\mapsto
M(\omega)\in\cB(l^2(\bZ\times\bN));\infty>\Vert
M\Vert_{\Omega,r,\delta}:=\\
&&\left.
\sum_{d\in\bZ^2}e^{\vert d\vert
r}\langle \vert d\vert\rangle^\delta
\sup_{\omega\neq\omega^\prime, i-j=d}\left(\vert
M_{ij}(\omega)\vert+\vert{M_{ij}(\omega)-M_{ij}(\omega^\prime)\over
\omega-\omega^\prime}\vert\right)\right\}.\eeqno
The result of the 	KAM algorithm we need here is:

\begin{theorem}\label{kam} Let $\tau\in(0,\infty)$
be large enough,
$\Omega=\lbrack\omega_-,\omega_+\rbrack\subset(0,\infty)$,
$M=M(\omega)=M^\ast(\omega)$ a family of matrix operators in
$l^2(\bZ\times\bN)$ such that
\begin{enumerate}
\item[] $\Vert{\bf O}M\Vert_{\Omega,0,\tau}<\infty$,
$M_{jj}=\omega 
j_1+e_{j_2}(\omega)\quad(j=(j_1,j_2)\in\bZ\times\bN)$ \hbox{with}
\item[] $\inf_{\omega,n}{e_{n+1}-e_{n}\over n}\ge cte>0$,
\item[] $\vert\vert\vert e\vert\vert\vert
:=\sup_{\omega\neq\omega^\prime,n,m}
\left\vert{(e_n-e_m)(\omega)-(e_n-e_m)(\omega^\prime)
\over\omega-\omega^\prime}\right\vert<1$.
\end{enumerate}
Then there is a $\delta\in (0,\tau)$ such that for $\gamma$ small
enough and $\Vert {\bf O}M\Vert_{\Omega,0,\tau}\le\gamma^2$ there is
a set of good frequencies $\Omega_\infty\subset\Omega$ with measure
\beqno \vert\Omega\setminus\Omega_\infty\vert=O(\gamma)\eeqno
and a unitary family $U_\infty(\omega)$ with $\Vert
U_\infty\Vert_{\Omega,0,\delta}<\infty$ such that
\beqno U_\infty
MU_\infty^{-1}(\omega)=M_\infty(\omega),\quad
{\bf O}M_\infty=0\qquad(\omega\in\Omega_\infty).\eeqno
\end{theorem}
\begin{remarks}
\item[(i)] In particular $M(\omega)$ has a basis of
eigenfunctions
$f_j$ which decay polynomially:
$f_j(k)={(U_\infty^{-1}})_{kj}=O(\vert k-j\vert^{-\delta})$;
\item[(ii)] Actually $\delta=\tau-cte$ so $\delta$ can be chosen
arbitrarily large if $\tau$ is arbitrarily large.
\end{remarks}

{\bf Outline of the Proof.}
The KAM method in its quantum guise first introduced by \cite{bell}
is by now quite standard. We shall, however, give only a descriptive
proof and refer to
\cite{duclstov} and references therein for analytic details. 

The idea is to successively diminish the
size of small off-diagonal elements by unitary transformations. We
use the function
$\chi(True):=1, \chi(False):=0$. Denote the matrices
\beqno{\bf D}M_{ij}:=M_{ij}\chi(i=j),\quad {\bf O}M:=M-{\bf D}M,\\
{\bf D}_nM_{ij}:=M_{ij}\chi(\vert i-j\vert=n),\quad {\bf
B}_nM:=\sum_{j=0}^n{\bf D}_nM,\eeqno
and define recursively
\beqno &&M_1:={\bf B}_1M,\quad U_0:=\bI,\\
&&W_{n;ij}:={M_{n;ij}\over M_{n;ii}-M_{n;jj}}\chi(i\neq j),\quad
U_n:=e^{W_n}U_{n-1},\\
&&M_{n+1}:=U_n({\bf
B}_{n+1}M)U_n^{-1}=e^{W_n}M_ne^{-W_n}+U_n({\bf
D}_{n+1}M)U_n^{-1}.\eeqno 
The idea of this is based on the identity
\beqno ad_{W_n}({\bf D}M_n):=\lbrack W_n,{\bf D}M_n\rbrack=-{\bf
O}M_n\eeqno
which by the Lie Schwinger formula
\beqno e^WMe^{-W}=\sum_{k=0}^\infty{ad_W^k(M)\over k!}\eeqno
leads to
\beqno M_{n+1}={\bf
D}M_n+\sum_{k=1}^\infty{k\over(k+1)!}ad^k_{W_n}({\bf O}M_n)+U_n{\bf
D}_{n+1}MU_n^{-1}.
\eeqno
$W_n$ is to be estimated by ${\bf O}M_{n}$ so the second term in
${\bf O}M_{n+1}$ will be quadratic in $\Vert{\bf O}M_n\Vert$. It is in
this estimate where one looses the resonant $\omega$ giving rise to
small divisors. For each step one proves that  for
$\sigma>1$ and $\gamma_n$ small enough there is an open set
$\Omega_{n+1}\subset\Omega_n$ with
$\vert\Omega_n\setminus\Omega_{n+1}\vert\le cte\
{\gamma_n\over1-\vert\vert\vert e \vert\vert\vert}$ such that for
$r_{n+1}<r_n$ it holds:
\beq\label{gamma}\Vert
W_n\Vert_{\Omega_{n+1},r_{n+1},\delta}\le{cte\over
\gamma_n^2(r_n-r_{n+1})^{2\sigma+1}}\Vert{\bf
O}M\Vert_{\Omega_n,r_n,\delta}\ ;\eeq
here $\vert\vert\vert e \vert\vert\vert$ estimates the space part
of ${\bf D}M_n$. The bad frequencies are controlled by a diophantine
estimate
\beqno
\Omega_n\setminus\Omega_{n+1}=\bigcup_{k,m,n\in\bN,n>m}\{\omega;
\vert\omega k+e_m-e_n\vert<\gamma( k+ n-m)^{-\sigma}\}.\eeqno
The growing gap property is then used to show that the
contributions to the measure are summable.
Estimating now $\Vert ad^k_WM\Vert\le cte^k\Vert W\Vert^k\Vert
M\Vert$ it follows for $r_{n+1}< r_n$ with the
shorthand $\Vert\cdot\Vert_n:=\Vert\cdot\Vert_{\Omega_n,r_n,\delta}$

\beqn\label{estimate} &&\Vert {\bf O}M_{n+1}\Vert_{n+1}\le
\\ &&cte\Vert W_n\Vert_{n+1} e^{cte \Vert W_n\Vert_{n+1}}\Vert{\bf
O}M_n\Vert_{n}+e^{2\sum_j \Vert W_j\Vert_{n+1}}\Vert{\bf
D}_{n+1}M\Vert_{n},\nonumber\eeqn
\beqno &&\Vert {\bf
D}M_{n+1}-{\bf D}M_{n}\Vert_{n+1}\le \\
&&cte\Vert W_n\Vert_{n+1} e^{cte \Vert W_n\Vert_{n+1}}\Vert{\bf
O}M_n\Vert_{n}+e^{2\sum_j \Vert W_j\Vert_{n+1}}\Vert{\bf
D}_{n+1}M\Vert_{n},\eeqno
\beqno &&\Vert U_n^{\pm}\Vert_{n+1}\le e^{\Vert W_n\Vert_{n+1}}
\Vert U_{n-1}^{\pm}\Vert_{n}.\eeqno

The choice  $\gamma_n=O(1/n^\mu), r_n=O(1/n^{\nu-1})$ for suitable
$\mu,\nu$ in estimate (\ref{estimate}) then leads to a quadratic
estimate for $\Vert W_n\Vert_{n+1}$:
\beqno\Vert W_n\Vert_{n+1}\le c_1\Vert W_n\Vert_{n+1}^2 e^{cte \Vert
W_n\Vert_{n+1}}+c_2\langle n\rangle^\beta\eeqno
where the constant $c_2$ is proportional to $\Vert{\bf
O}M\Vert_{\Omega,0,\tau}$, $\beta=2\mu+\nu(2\sigma+1)-\tau$. If
$\Vert{\bf O}M\Vert_{\Omega,0,\tau}$ is small enough and $\tau$ large
enough this implies that $\Vert W_n\Vert$ is summable, and that
${\bf D}M_n$ and $U_n$ are convergent.
\hfill
$\Box$\\[2mm]
As a consequence of this and the analysis in section \ref{matrix} we
obtain that the spectrum is pure point:
\begin{corollary}\label{purepoint}
For the Floquet Hamiltonian defined in equation (\ref{K}) it holds:
\beqno K(\omega,g) \hbox{ has a basis of eigenvectors in }
L^2((0,T),dt;L^2((0,L),dx))\eeqno
provided $g$ is small enough and
$\omega\in\Omega_\infty\subset\lbrack\omega_-,\omega_+\rbrack$ the set
constructed in Theorem (\ref{kam}) with measure
$\vert\lbrack\omega_-,\omega_+\rbrack\setminus\Omega_\infty\vert=
O(\sqrt{g})$.

\end{corollary}

{\bf Proof.} Identifying $L^2((0,L),dx)$ with $l^2(\bN)$ via the
eigenbasis $\{\psi_n\}$ of $H$ constructed in Theorem \ref{spectrum}
we find that $K$ is unitarily equivalent to
\beqno D_t+E_n(t)\delta_{nm}+\langle\psi_n, D_t\psi_m\rangle\eeqno
on $L^2((0,T),dt;l^2(\bN))$, which is of the form
\beqno D_t+h_0(\omega,g)+gV(t,\omega,g)\eeqno
where $V$ is $C^\infty$ bounded and $O(1)$ in $g$, and
$h_0(\omega,g)_{ij}:=\langle E_j\rangle\delta_{ij}$. Applying a
version of the superadiabatic regularization as in
\cite{howl},\cite[Thm.3.6]{duclstov} we get that $K$ is
unitarily equivalent to an operator of the same form whose
fluctuating part has the property that
$(m^2-n^2)^\tau V_{mn}$ is $C^\infty$ bounded in $l^2$ uniformly in
the parameters for any $\tau>0$. Furthermore, by Theorem \ref{H(t)} the
diagonal elements still satisfy
${e_{n+1}-e_n\over n}\ge cte>0$ and $0\le\partial_\omega e_n<1$ for
$g$ small enough. 

Finally going to the Fourier representation $K$ turns out to be
unitarily equivalent to a matrix $M$ on $l^2(\bZ\times\bN)$ which
has the properties required in Theorem \ref{kam}.
\hfill
$\Box$\\[2mm]

By pure pointness of $\sigma(K)$ every trajectory
$\{t\in\bR;U(t)\psi\}$ generated by $H$ is a precompact set. It
follows that
\beqno
\lim_{r\to\infty}\sup_{t\in\bR}\Vert\chi(H(t)>r)U(t)\psi\Vert=0.\eeqno
This does, however, not imply that the energy expectation
$\vert\langle\psi,H(t)\psi\rangle\vert$ is bounded, as the example
given in
\cite{delrjitolastsimo2} shows.

We shall show now that boundedness of the energy is in fact always
ensured in cases where the KAM algorithm used above applies.

\begin{corollary}\label{energygrowth}
For $g$ small enough there exists a  set of frequencies
$\Omega_\infty\subset\lbrack\omega_-,\omega_+\rbrack\subset(0,\infty)$
with
$\vert\lbrack\omega_-,\omega_+\rbrack\setminus\Omega_\infty\vert=
O(\sqrt{g})$
such that for $\omega\in\Omega_\infty$ and for the propagator $U$ of
$H(t,\omega,g)$ it holds
\begin{enumerate}
\item[(1)] 
$ U(t)=U_p(t)e^{-iGt}U_p^{-1}(0)$

where $G$ commutes and is relatively bounded with respect to 
$H(0)$  and
 $U_p$  is $T$ periodic  and  $C^\infty$  as a bounded
operator  in  
$L^2((0,L),dx)$;
\item[(2)] 
$ \vert \langle U(t)\psi, H(t) U(t)\psi\rangle\vert\le
cte$

for $\psi\in Q(H(0))$, uniformly for $t\in\bR$.

\end{enumerate}
\end{corollary}
{\bf Proof.} Application of the KAM  algorithm to the matrix $M$ of
$K$ gave $U_\infty MU_\infty^{-1}=M_\infty$. We transform back to the
space of $(t,x)$ functions using the basis $\{\phi_j\}$
whose space part is $\{\psi_{j_2}\}$ as constructed in Theorem
\ref{spectrum}. Denote
\beqno \cT_\infty:=\sum_{j,k\in\bZ\times\bN}
({U_\infty})_{jk}\vert\phi_j\rangle\langle\phi_k
\vert.\eeqno
By construction
${U_\infty}$  is a Toeplitz matrix in the indices corresponding to
the Fourier variable, i.e.
$({U_\infty})_{jk}$ is of the form $({U_\infty})_{j_1-k_1,j_2,k_2}$.
Consequently 
$\cT_\infty$ is fibered, i.e.:
$\cT_\infty\psi(t,x)=T_\infty(t)\psi(t,x)$ for $T_\infty(t)$
unitary, periodic and $C^\infty$ bounded in $L^2((0,L),dx)$;
$(M_\infty)_{jj}=\omega j_1+e^\infty_{j_2}$ with
$e^\infty_{j_2}-\langle E_{j_2}\rangle=O(g)$ uniformly in
$n,\omega$. The reader may consult
\cite{duclstov} for a more detailed discussion of this point. By
Theorem \ref{kam} this results in
\beqno \cT_\infty K\cT_\infty^{-1}={\sum
(M_\infty})_{jj}\vert\phi_j\rangle\langle\phi_j\vert=D_t+H_\infty(t)
\eeqno
where $H_\infty$ is defined by $(D_t+H_\infty(t))\psi_m(t)=e^\infty_m
\psi_m(t)$. Denote
\beqno
U_A(t):=\sum_m\vert\psi_m(t)\rangle\langle\psi_m(0)\vert\eeqno
then the relation 
\beqno(D_t+H_\infty(t))U_A(t)=U_A(t)\sum e^\infty_mP_m(0)\eeqno
holds on  $D(H(0))$ so with the definition $G:=\sum
e^\infty_mP_m(0)$ we obtain 
\beqno (U_A^{-1}T_\infty) H(T_\infty^{-1}U_A)+
\left((U_A^{-1}T_\infty) D_t(T^{-1}_\infty U_A)\right)
=G.\eeqno

This formula implies the asserted
form for the propagator  with the definition
$U_p(t):=T_\infty^{-1}U_A(t)$ and the fact that 
$U_A$ is $C^\infty$ bounded by Theorem
\ref{spectrum}, furthermore it shows that $U_p$ preserves domains:
$U_p(t)D(H(0))\subset D(H(t))$.

For $\psi$ in the form domain of $H(0)$ it holds with
$\varphi:=U_p^{-1}(0)\psi$ 
\beqno &&\langle U(t)\psi,
H(t)U(t)\psi\rangle=-\langle U(t)\psi,
D_tU(t)\psi\rangle\\
&&=\langle
e^{-iGt}\varphi,(G-(U_p^{-1}(D_tU_p))(t)e^{-iGt}\varphi\rangle\\
&&=\langle\varphi, G\varphi\rangle-
\langle\varphi,e^{iGt}(U_p^{-1}(D_tU_p))(t)e^{-iGt}\varphi\rangle \eeqno
which is bounded uniformly in time as 
$(U_p^{-1}(D_tU_p))$ is  periodic. \hfill
$\Box$\\[2mm]

By the same method we complement now the results of
\cite{nenc} which were much extended in 
\cite{joye,barbjoye}. These authors estimate the propagator by time
dependent methods to discuss stability of the energy expectations. The
spectral method used here is better suited to provide  bounds valid on
an infinite time scale. The differentiability properties on the
potential could be relaxed. However, we do not make an effort, here,
to do so.

\begin{theorem}\label{generalenergygrowth}
Let $T,g>0$, $W$ a $T$ periodic $C^\infty$ function with values in
the bounded operators on a Hilbert space. Consider $H_0=\sum_n
E_nP_n$ with  growing gaps: ${E_{n+1}-E_n\over n^\alpha}\ge cte>0$
for an
$\alpha>0$.

Then it holds for the propagator $U$ of $H(t):=H_0+gW(t)$, $\psi\in
Q(H_0)$:
\beqno \vert \langle U(t)\psi, H(t) U(t)\psi\rangle\vert\le
cte\eeqno
provided $g$ is small enough and
$\omega\in\Omega_\infty\subset\lbrack\omega_-,\omega_+\rbrack\subset
(0,\infty)$,
the set constructed as in Theorem (\ref{kam}) with measure
$\vert\lbrack\omega_-,\omega_+\rbrack\setminus\Omega_\infty\vert=
O(\sqrt{g})$.
\end{theorem}
{\bf Proof.} By \cite{duclstov} the KAM algorithm is applicable,
so we can proceed as in the previous corollary and find
$U_p,G$ with $U(t)=U_p(t)e^{-iGt}U_p^{-1}(0)$. 
\hfill
$\Box$\\[2mm]

\section{Stability for resonant frequencies}\label{resonant}In this
section we shall show that for resonant frequencies
$\omega\in\bQ({L\over\pi})^2$ the spectrum of $K$ is still pure
point.

It is equivalent to show that the time $T$ map $U(T)$ of $H$ has pure
point spectrum. We go to the matrix representation of section
\ref{matrix}.  Let $\{\psi_n\}$ be the basis found in Theorem
\ref{spectrum},
$\{e_n\}$ the standard basis of $l^2$ and $J=J(t,\omega,g)$ the
unitary operator $J:=\sum_0^\infty\vert
e_n\rangle\langle\psi_n\vert$. We have
\beqno J(t)(D_t+H(t))J^{-1}(t)=D_t+h(t)\\
h_{nm}:=E_n\delta_{nm}+\langle\psi_n,D_t\psi_m\rangle.\eeqno
An approximation  of $E_n$ was worked out in equation
(\ref{en}). Let $g_n(t)=g_n(t+T):=\int_0^t{4g\over
L}(-1)^{n+1}\cos{\omega s}\ ds$, $G$ the gauge transformation
defined by $G_{nm}(t)=\exp(i g_n(t))\delta_{nm}$. Now 
\beqno &&G(D_t+h)G^{-1}-\left(D_t+(({n\pi\over L})^2+{\omega\over2}+
\langle W\rangle+{4g\over L})\delta_{nm}\right)\\
&&=O({1\over
n})\delta_{nm}+e^{i(g_n-g_m)}\langle\psi_n,D_t\psi_m\rangle;\eeqno 
by Theorem \ref{spectrum}  the matrix function on the right
hand side is a $C^\infty$ function in the   
Hilbert Schmidt norm $\Vert a\Vert_{HS}:=(\sum_{nm}\vert
a_{nm}\vert^2)^{1/2}$ and a forteriori in the compact operators on
$l^2(\bN)$.  We now make use of the argument of En\ss\ and Veseli\'c
to conclude: \begin{theorem} For $\omega\in\bQ({L\over\pi})^2, g$
small enough, $T={2\pi\over\omega}$ it holds \begin{enumerate}
\item[(i)] $\sigma_{ess}(U(T,\omega,g))= \left\{\exp({-i(({\pi n\over
L})^2+{\omega\over2}+\langle W\rangle+{4g\over
L})T}); n\in\bN\right\}$, 
\item[(ii)] $L^2((0,L))$ has a basis of eigenvectors of
$U(T,\omega,g)$.
\end{enumerate}
\end{theorem}
{\bf Proof.} Denote $\tilde h:=GhG^{-1}+G(D_tG^{-1})$, 
$\tilde{h}_0:=(({n\pi\over L})^2+{\omega\over2}+
\langle W\rangle+{4g\over L})\chi(n=m)$ and by $\tilde U,\tilde U_0$
their propagators. The spectrum
\beqno\sigma(\tilde U_0(T,\omega,g))=\left\{\exp({-i(({\pi n\over
L})^2+{\omega\over2}+\langle W\rangle+{4g\over
L})T}),n\in\bN\right\}\eeqno
is a discrete set. Furthermore 
\beqno\int_s^t \tilde U_0^{-1}(\tilde h-\tilde h_0)\tilde U_0\eeqno
is compact for every $s,t\in\bR$. By Theorem 5.2 of \cite{enssvese}
$\tilde U(T)-\tilde U_0(T)$ is compact. So
$\sigma_{ess}(\tilde U(T))=\sigma_{ess}(\tilde U_0(T))$, which cannot
contain continuous spectrum so $\sigma(\tilde U(T))$ is pure point.
$G(T)=\bI$, from the unitary equivalence
\beqno U(T)=J^{-1}(T)\tilde U(T)J(T).\eeqno
we  conclude that the spectrum of $U(T)$ is pure point 
\hfill
$\Box$\\[2mm]

\subsection*{Acknowledgments}
Part of the results announced here was obtained during the visits of
J.A. and P.D. to the Nuclear Physics Institute AS CR, \v Re\v z, and P.E. to
the Universit\'e de Toulon et du Var and CPT CNRS, Marseille; the
authors express their gratitude to the hosts. The work was partially supported
by the AS CR Grant No. 1048801

\end{document}